\documentclass[amsmath, preprintnumbers, 10pt, twocolumn, pre, bibliograpy]{revtex4-1}

\usepackage{graphicx}
\usepackage{subfigure}
\usepackage{color}
\usepackage{amsfonts}
\usepackage{amsmath}
\usepackage{bbm}

\newcommand{\MA}{\mathbf A}
\newcommand{\MB}{\mathbf B}
\newcommand{\MX}{\mathbf X}
\newcommand{\aff}{\mathcal A}
\newcommand{\R}{\mathcal R}

\renewcommand{\Re}{{\mathrm Re}}
\renewcommand{\Im}{{\mathrm Im}}
\usepackage{physics}

\begin{document}
\title{Robust oscillations in multi-cyclic Markov state models of biochemical clocks}
\author{Clara \surname{del Junco} and Suriyanarayanan Vaikuntanathan}
\email{Corresponding Author: svaikunt@uchicago.edu}
\affiliation{Department of Chemistry and The James Franck Institute, University of Chicago, Chicago, IL, 60637}

\begin{abstract}
Organisms often use cyclic changes in the concentrations of chemicals species to precisely time biological functions. Underlying these biochemical clocks are chemical reactions and transport processes, which are inherently stochastic. Understanding the physical basis for robust biochemical oscillations in the presence of fluctuations has thus emerged as an important problem. In a previous paper [C. del Junco and S. Vaikuntanathan, Phys. Rev. E {\bf 101}, 012410 (2020)], we explored this question using the non-equilibrium statistical mechanics of single-ring Markov state models of biochemical networks that support oscillations. Our finding was that they can exploit non-equilibrium driving to robustly maintain the period and coherence of oscillations in the presence of randomness in the rates. Here, we extend our work to Markov state models consisting of a large cycle decorated with multiple small cycles. These additional cycles are intended to represent alternate pathways that the oscillator may take as it fluctuates about its average path. Combining a mapping to single-cycle networks based on first passage time distributions with our previously developed theory, we are able to make analytical predictions for the period and coherence of oscillations in these networks. One implication of our predictions is that a high energy budget can make different network topologies and arrangements of rates degenerate as far as the period and coherence of oscillations is concerned. Excellent agreement between analytical and numerical results confirms that this is the case. Our results suggest that biochemical oscillators can be more robust to fluctuations in the path of the oscillator when they have a high energy budget.
\end{abstract}

\maketitle 

\section{Introduction} 

Many organisms use internal biochemical clocks to synchronize their metabolism to day-night cycles, a tactic that confers fitness as it allows them to anticipate periodic environmental changes~\cite{Woelfle2004}. These clocks are implemented as a series of chemical reactions and transport processes, whose timing can be affected by intrinsic and extrinsic noise - yet they have evolved to maintain consistent periods over different copies of the oscillator (e.g. in different cells), and over time. For example, the circadian oscillator of {\it S. Elongatus} bacteria can be reconstituted {\it in vitro} from just 3 proteins, called KaiA, B, and C, which can sustain oscillations in the phosphorylation level of KaiC with a 24-hour period over many days even in constant light or dark conditions~\cite{Nakajima2005, Tomita2005, Rust2007}.  From a theoretical standpoint, understanding the non-equilibrium statistical mechanical requirements for maintaining robust oscillations in these molecular clocks has thus emerged as an important question. In particular, the positive connection between the amount of energy dissipated and the precision of the stochastic period of these oscillators has been noted by many theoretical studies~\cite{Barato2015, Cao2015, Barato2017, Fei2018, Wierenga2018, Nguyen2018, Marsland2019}. Some of this work proposes thermodynamic bounds which set a lower limit on the extent of stochastic fluctuations in these systems as a function of the energy dissipation budget~\cite{Barato2015, Barato2017, Wierenga2018}.  However, the structure of even simple models of biochemical oscillators constrains them to operate far from these bounds~\cite{Marsland2019}, which raises the question of what role energy dissipation plays in these cases where fluctuations are much larger than the minimum for the amount of energy that the oscillator is using.

\begin{figure}
\centering
\includegraphics[width=\linewidth]{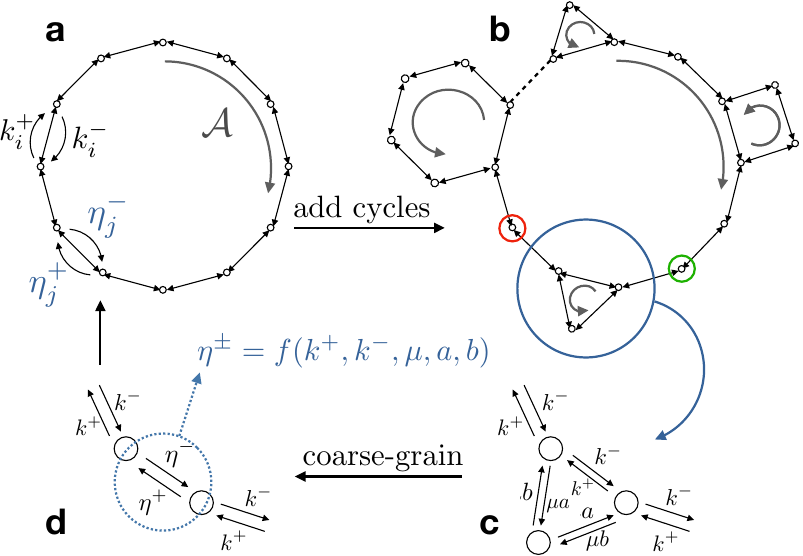}
 \caption{A schematic of the networks studied in this article. (a) In previous work~\cite{DelJunco2018b}, we developed an analytical theory for the period of oscillations $T$ and the number of coherent oscillations $\R$, defined in Eq.~\ref{eq:r-t}, in oscillators that can be represented by a single cycle of states where the clockwise (CW) hopping rates ($k_i^+$) are much larger than the counterclockwise (CCW) hopping rates ($k_i^-$). This asymmetry is quantified by the affinity $\mathcal A$, defined in Eq.~\ref{eq:aff}. (b) In this work, we extend our results to networks where the main cycle is decorated with many small cycles. (c) We design these `decorations' so that the rates going into them are modulated by a small parameter $\mu$ which governs the probability that the system will enter the decoration. (d) To apply our theory to these multi-cycle networks, we map the muti-cyclic network on to a unicyclic network by matching moments of the first passage time distribution for a random walk beginning CCW from the decoration (green circle) and ending CW from it (red circle) on to a line of states with two unknown hopping rates. This procedure gives a set of effective hopping rates $\eta_i^\pm$. When each of the decorations in the network is coarse-grained in this manner, we obtain a single-cycle network as in (a) whose period $T$ and coherence $\R$ approximate those of the full, multi-cycle network.}
\label{fig:schematic}
\end{figure}

Recently, we explored this question using single-cycle Markov models of biochemical oscillators such as that pictured in Fig.~\ref{fig:schematic}a~\cite{DelJunco2018b}. By deriving an analytical expression for the period and coherence of oscillations that reveals their detailed dependence on all of the rates in the network, we showed that non-equilibrium driving allows the period of oscillators to become insensitive to many of the parameters of the models - specifically, the arrangement of the transition rates on the ring. Driving thus allows the period of a wide class of oscillators - even those operating far from the bound - to be robustly maintained in the presence of changes in these parameters. In this paper, we further explore this role of energy dissipation by extending our results to networks with multiple cycles. We show that the period of the oscillator is more robust at high driving to changes in the the topology of the network as well as the rates.

The paper is organized as follows: in Section~\ref{model} we introduce the class of multi-cyclic Markov state models considered in this paper and define non-equilibrium driving, the observables of interest (period and coherence of oscillations), and robustness in the context of these models. In Section \ref{theory} we briefly review our analytical theory from Ref.~\citenum{DelJunco2018b}. This theory depends on the single-cycle topology of the networks studied in Ref.~\citenum{DelJunco2018b}, so to apply it to multi-cycle networks such as the one illustrated in Fig.~\ref{fig:schematic}b, in Section~\ref{CG} we show how to coarse-grain small cycles, which we call ``decorations", on to single links, yielding an effective single-cycle network whose period and coherence are meant to approximate those of the full multi-cyclic network. In Section~\ref{timescales} we compare these observables calculated numerically in multi-cycle networks to the corresponding coarse-grained networks and analytical approximations, and show that at high affinity our analytical approximation, which takes as input only a small subset of the parameters required to specify the multi-cycle network, accurately reproduces the period and coherence of these networks. We demonstrate the ability of our theory to predict timescales when the rates and topology in the network are randomly generated. Finally, in Section~\ref{disc}, we discuss the implications of our results for biochemical oscillators and show one example of how the multi-cycle networks studied here can be used to achieve input compensation, which is the ability to maintain a constant period when the affinity changes.

\section{Markov state models of biochemical oscillators}\label{model}

In this paper we consider Markov models, such as the ones in Fig.~\ref{fig:schematic}b, as simple models that capture the cycling and stochasticity of biochemical oscillators~\cite{Barato2017}.  In these models, each vertex represents a collective state of the system. For instance, in the KaiABC oscillator, it could be a vector of the counts of KaiC monomers in each phosphorylation state, and the concentrations of other species in solution~\cite{Rust2007}. The rates along each of the edges represent the rates of elementary processes, like a phosphorylation event. We emphasize that this picture is thus not a representation of the underlying chemical reaction network which must contain, at a minimum, a negative feedback loop, and may also have other motifs~\cite{Novak2008}. Rather, it is an emergent picture that captures the oscillations that can arise from such a network, and the feedback as well as mass action kinetics are encoded in the rates along each edge, which depend on the collective state of the system represented by the connected vertices. It is not expected that a real oscillator will always follow the same path through its state space on each cycle. The single-cycle model in Fig.~\ref{fig:schematic}a is a caricature that captures the average limit cycle of the oscillator. The multi-cycle model in Fig.~\ref{fig:schematic}b is a caricature intended to reflect small fluctuations about this average path. 

The network is driven out of equilibrium by an affinity, $\mathcal A$, defined as 
\begin{equation}
\aff = \sum_{cycle} k_i^{+}/k_i^{-}.
\label{eq:aff}
\end{equation}
A finite affinity is necessary to have oscillations. The affinity is formally defined on closed cycles. In this work we refer to the affinity per site, $\aff/N = \langle k_i^{+}/k_i^{-} \rangle_{cycle}$, as a measure of the strength of driving. In biology, this non-equilibrium driving is typically provided by ATP hydrolysis and the affinity over the cycle will quantify the total net turnover of ATP per cycle. The quantities of interest are two time scales: the average period of oscillations $T$ and the number of coherent oscillations, $\R$, defined as
\begin{align}
T = 2\pi/ |\Im[\phi]| && \R = -|\Im[\phi]|/\Re[\phi] 
\label{eq:r-t}
\end{align}
where $\phi$ is the eigenvalue of the transition rate matrix of the network that yields the largest value of $\R$~\footnote{In refs.~\citenum{DelJunco2018b} and \citenum{Barato2017}, $\phi$ was defined as the eigenvalue with the least negative real part. However, in the multi-cyclic networks we address later in this paper, that definition can lead to selecting an eigenvalue which corresponds to cycling around a small decoration rather than global oscillations.}. Loosely speaking, a higher value of $\R$ corresponds to smaller fluctuations in the period. Although $\R$ is one measure of the quality of timekeeping in the clock, it is not the definition of robustness that we use in this paper. In \citenum{Barato2017}, it was postulated that $\mathcal R$ is maximized in a uniform oscillator, that is, when all of the clockwise (CW) rates $k^+_i$ in the model in Fig.~\ref{fig:schematic}a are equal to one another and related to the counterclockwise (CCW) rates $k^-_i$ by $k^+ = \exp(\aff/N) k^-$. However, since the rates along each edge of the network depend on the collective state of the system represented by the connected vertices, a network that sustains oscillations cannot be uniform~\cite{Marsland2019}. Moreover, the rates and, in the case of the multi-cyclic network depicted in Fig.\ref{fig:schematic}b, the locations and sizes of the small secondary cycles, can fluctuate over time and between copies of the oscillator. In this paper we therefore consider two properties of the oscillator: first, how predictable the period of oscillations $T$ and the coherence $\R$ are with limited knowledge about the specific details of the arrangement of rates and decorations in the network, and by extension, how robust these quantities are to fluctuations in these details.

\section{Analytical theory for timescales in single-cycle oscillator models}\label{theory} 

In Ref.~\citenum{DelJunco2018b}, we derived analytical expressions for $\phi$, and therefore the period and coherence of oscillations, in a network consisting of a single cycle of states, as depicted in Fig.~\ref{fig:schematic}a. In this section, we briefly review that result. Further details are available in Ref.~\citenum{DelJunco2018b}. An exact expression for $\phi$ in a cycle of $N$ states in the special uniform case where $k_i^{+} = k^+$ and $k_i^{-} = k^-$ for all $i$ is given by:
\begin{equation}
 \phi^{(0)} = -(k^- + k^+) + k^-e^{-2\pi i/N} + k^+e^{2\pi i/N} \label{eq:phi0}.
\end{equation}
We then considered networks where at least one of the CW rates is equal to $k^+$ and at least one of the CCW rates is equal to $k^-$. The remaining $m \leq N-1$ rates, denoted $h_j^\pm$, can be assigned arbitrary values ranging at least an order of magnitude above or below these ``uniform" rates. The main result of Ref.~\citenum{DelJunco2018b} was an expression for $\phi$ in this setup, in the limit of high affinity where terms of order $k^-/k^+ = \exp(-\aff/N)$ can be neglected compared to terms of order 1. The result is summarized in the following expressions: 
\begin{widetext}
\begin{align}
\phi &= \phi^{(0)} + C \gamma \label{eq:phinew} \\
 \phi^{(0)} & = -(k^- + k^+) + k^-\exp(-2 \pi i/N) + k^+\exp(2 \pi i/N) \\
\gamma &= \frac{1}{m-N}\left(\sum_{j=1}^m \log(\zeta_j(\gamma, k^\pm, h_j^\pm, N))\right) + \frac{1}{2(m-N)^2}\left(\sum_{j=1}^m \log(\zeta_j(\gamma, k^\pm, h_j^\pm, N))\right)^2 \label{eq:gamma} \\
\end{align}
\end{widetext}
with expressions for $\zeta_j(\gamma, k^\pm, h_j^\pm, N))$ and $C$ given in Appendix~\ref{app:theory}. The essential feature of these equations is that Eq.~\ref{eq:gamma} depends independently on each rate $h_j^\pm$ and does not contain any information about the relative positions of the rates in the network. As a result, in the limit of moderately high affinity, spatial correlations vanish and the rates only contribute additively to the timescales, so we find that $\R$ and $T$ are insensitive to the arrangement of the rates in the network.  From a biological perspective, this means that the farther an oscillator operates from equilibrium, the more robust it will be to changes in the relative position of the rates. In a large enough network these rearrangements are akin to a scenario in which the values of the rates change but the distribution from which they are drawn stays the same (i.e., the rates fluctuate), and hence the oscillator is also more robust to fluctuations in the rates.

Since Eq.~\ref{eq:gamma} is a self-consistent equation for $\gamma$, the theoretical predictions in Section~\ref{timescales} are obtained numerically by searching for solutions to Eq.~\ref{eq:gamma} near to a linear approximation of Eq.~\ref{eq:gamma}.

\section{Mapping multi-cycle to single-cycle oscillators via first passage time distributions}\label{CG} 

We now wish to apply Eqs.~\ref{eq:phinew} - \ref{eq:gamma} to the networks with multiple cycles depicted in Fig.~\ref{fig:schematic}b in order to understand how our conclusions extend to these higher-dimensional cases. The derivation of Eqs.~\ref{eq:phinew} - \ref{eq:gamma} used a transfer matrix technique which depended on the single-ring topology of the network. Rather than trying to extend this approach to networks with decorations of the kind we wish to consider here, depicted in Fig.~\ref{fig:schematic}a, we took a different approach and chose instead to map multi-cycle networks on to single-cycle networks so that Eqs.~\ref{eq:phinew}-\ref{eq:gamma} can then be directly applied to the mapped network. Because we want a mapping that preserves time scales, our approach is to build a single-cycle network with rates such that the mean and variance of the first passage time from a site upstream (in the sense of the probability current) of a decoration to a site downstream of the decoration is preserved (denoted by the green and red circles in Fig.~\ref{fig:schematic}b). For each decoration, we replace the rates along the edge shared by the large and small cycles with an effective CW rate $\eta^+$ and an effective CCW rate $\eta^-$. 

In order to do this, we first calculate the first passage time (FPT) distribution across the decoration (from the green circle to the red circle in Fig.~\ref{fig:schematic}b) in Laplace space~\cite{Murugan2012, Budnar2019} (details in Appendix \ref{app:fpt}).The Laplace-transformed FPT distribution $\tilde F(s)$ is a moment-generating function for $F(t)$, with the $n$th moment $\langle \tau^n \rangle$ given by:
\begin{equation}
(-1)^n\left(\frac{\partial^n \hat F}{\partial s^n}\right)_{s = 0} = \langle \tau^n \rangle.
\label{eq:genfunc}
\end{equation}
Moments of the FPT distribution can thus be computed even when it is not easy to invert $\tilde F(s)$ to obtain the real-time FPT distribution $F(t)$. For the decoration in Fig.~\ref{fig:schematic}c, $\tilde F_{dec}(s)$ is a function of $\mu, a, b, k^{+}$, and $k^{-}$. For the line of states in Fig.~\ref{fig:schematic}d,  $\tilde F_{line}(s)$ is a function of $k^{+}, k^{-}$, and two unknown rates $\eta^{+}, \eta^{-}$. By setting the mean and variance of $F_{dec}$ and $F_{line}$ equal to one another, we obtain analytical expressions for $\eta^{+}, \eta^{-}$ in terms of $\mu, a, b, k^{+}$, and $k^{-}$. These expressions are algebraically complicated, so we do not reproduce them here; for the smallest motif considered - a triangle as depicted in Fig.~\ref{fig:schematic}c - the full effective rates are given as an example in Appendix \ref{app:eff-rates}.  By calculating effective rates for all of the decorations in a network, we construct a single-cycle network that we expect to have a similar period $T$ (roughly captured by the first moment of the FPT distribution) and coherence $\R$ (roughly captured by the second moment of the FPT distribution) as the decorated network.

\begin{figure}
\centering
\includegraphics[width= \linewidth]{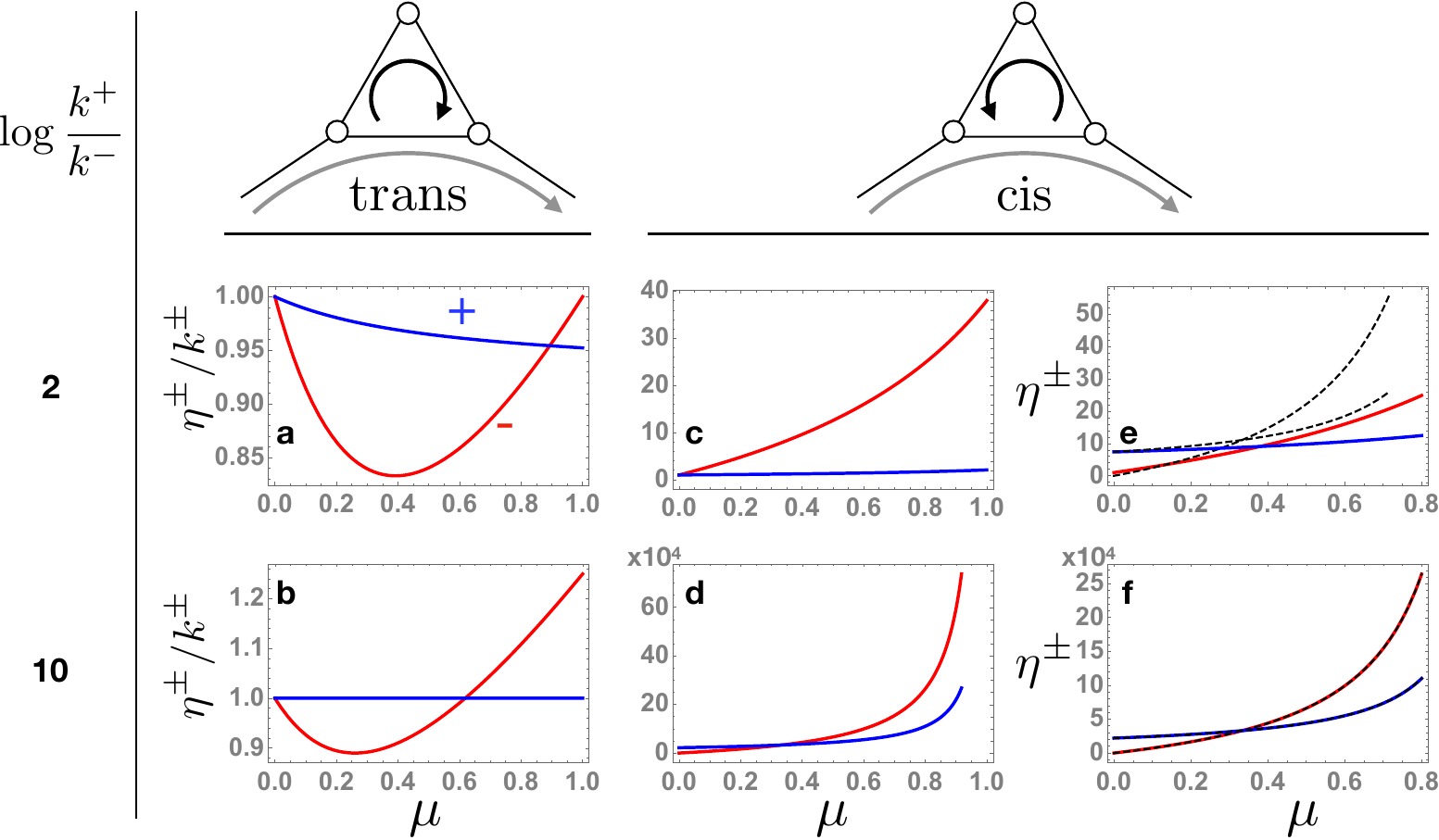}
 \caption{Effective rates resulting from the coarse-graining procedure as a function of $\mu$, which controls the probability of entering the triangle. We consider the cases where $a = k^+$ and $b = k^-$ ({\it ``trans"}, a-b) and where $b = k^+$ and $a = k^-$ ({\it ``cis"}, c-f).  a-d: relative values of the effective rates $\eta^\pm /k^\pm$ compared to the $\mu = 0$ values. The rates change much more dramatically in the {\it cis} case, because an extra, likely path for hopping CCW through the decoration is added. e-f: For the {\it cis} case, the absolute value of the effective rates and comparison with the high-affinity expressions given in table \ref{tab:rates} (black dashed lines).}
\label{fig:effrates}
\end{figure}

We note that this procedure does not always produce reasonable coarse-grained rates. The effective rates diverge at a value of $\mu$ that decreases as the size of the decoration increases (Table~\ref{tab:rates}). As the size of the decoration becomes larger, it will be able to support coherent oscillations of its own, leading to a system with multiple interacting periods of oscillation; in that case, one cycle is no longer dominant in terms of the dynamics of the system and we do not expect to be able to simply coarse-grain out these competing cycles. The effective rates can also become negative if the rates in the main cycle ($k^{+/-}$) and the decoration ($a/b$) are very (orders of magnitude) different. We therefore restrict our study to small cycles with 6 sides or fewer where the effective rates are positive for values of $\mu$ of at least $\mu = 0.05$, and to cases where $a = k^{-}$ and $b =  k^{+}$, which we call ``cis" because the rates in the main cycle and decoration favor current in the same direction through their shared edge, or where $a = k^{+}$ and $b =  k^{-}$, which we call ``trans" because the rates in the main cycle and decoration favor current in opposite direction through their shared edge. The probability of entering the decorations is tuned by changing $\mu$. The {\it cis} configuration favors cycling in the small decoration compared to the {\it trans} configuration, as shown schematically in Fig.~\ref{fig:effrates}. In Table~\ref{tab:rates} we show the effective rates in the {\it cis} configuration in the limit where $k^{-}/k^{+} \to 0$ (for the {\it trans} configuration this limit simple gives $\eta^+ = k^+$ and $\eta^- = 0$). 
 In Fig.~\ref{fig:effrates} we show $\eta^\pm$ as a function of $\mu$ for both configurations. The {\it cis} configuration leads to much more dramatic changes in the effective rates than the {\it trans} configuration (Fig.~\ref{fig:effrates}), and $\R$ and $T$ change by only a small fraction as $\mu$ in turned on in networks with {\it trans}-configured decorations. For the results in the following sections we therefore focus on decorations with {\it cis} rates.

\begin{table}
\centering
\begin{tabular}{|c|c|c|c|} \hline
Shape &  Exclusive Vertices  & $\eta^+$ & $\eta^-$ \\ \hline
triangle & 1 &  $k^+/(1-\mu)$ & $3k^+\mu /(1-\mu) $\\ \hline
square & 2 & $k^+/(1-3\mu)$ & $6k^+\mu /(1-3\mu)$ \\ \hline
pentagon & 3 & $k^+/(1-6\mu)$ & $10k^+\mu /(1-6\mu) $\\ \hline
hexagon & 4 & $k^+/(1-10\mu)$ & $15k^+\mu /(1-10\mu)$\\ \hline
general & x & $\frac{2k^+}{2 - x(x+1)\mu}$ &$ \frac{k^+(x+1)(1 + x/2)}{2 - x(x+1)\mu}$ \\ \hline
\end{tabular}
\caption{Effective rates across decorations with increasing number of sides, in the limit of $k^-/k^+ \to 0$. We deduce by inspection that the effective rates for a decoration with $x$ vertices that do not belong to the large cycle are: $\eta^+ = k^+/(1 - \alpha \mu)$, $\eta^- = (x + 1 + \alpha) k^+/(1 -\alpha \mu) $ where $\alpha = \sum_{i=1}^x i = x (x +1)/2$.}
\label{tab:rates}
\end{table}

\section{Predicting timescales in multi-cyclic networks}\label{timescales}

We can now compare $T$ and $\R$ for our coarse-grained networks to the full networks.  We calculate $T$ and $\R$ in three ways: first, by numerically diagonalizing the transition rate matrix of a network with explicit decorations ($\R/T_{exact}$); second, by numerically diagonalizing the transition rate matrix of the corresponding coarse-grained network ($\R/T_{CG}$), and third, using the theoretical expressions in Eqs.~\ref{eq:phinew} - \ref{eq:gamma} with the rates in the coarse-grained network as input ($\R/T_{th}$).

In the following sections we test the ability of our coarse-graining scheme combined with our analytical theory to predict $T$ and $\R$ in networks with increasing amounts of randomness. Since our theory does not contain information about the locations of the effective rates in the network, if we are able to predict these observables it means that they are insensitive to the locations of decorations on the network and will be robust to any changes in the locations of the decorations.  

\subsection{Networks with symmetrically distributed decorations}

\begin{figure}
\centering
\includegraphics[width=\linewidth]{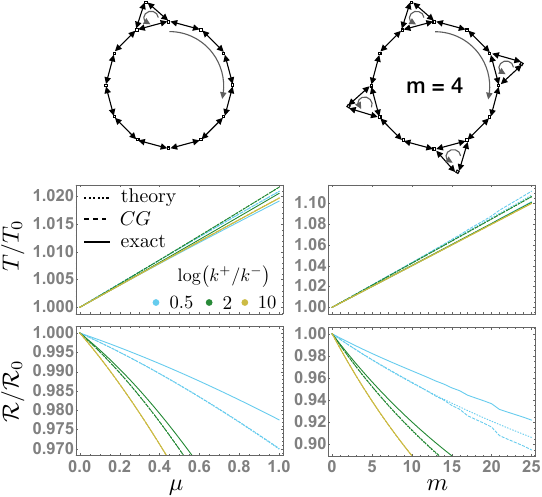}
 \caption{Period $T$ and number of coherent oscillations $\R$ in networks with $N = 100$ states in the main cycle and triangle decorations with the {\it cis} configuration of rates ($a = k^-, b = k^+$). We set $k^- = 1$ and set $k^+ = \exp(\aff_0/N)$. $\aff_0$, $R_0$ and $T_0$ are the respective values in a network with no decorations. ``Exact" results are calculated from numerical diagonalization of the full network. ``CG" results are calculated from numerical diagonalization of a single cycle with $N = 100$ states with effective rates through the links where decorations are located in the full network. ``Theory" results are calculated from the theoretical expression in Eqs.~\ref{eq:phinew} and \ref{eq:c1sq} with the rates $\{h_j^\pm\}$ given by the effective rates. On the left we show results for networks with one triangle decoration, as a function of $\mu$, which governs the probability of entering the decoration (see Fig.~\ref{fig:schematic}). On the right, we show results for networks with $\mu = 0.2$ fixed and with $m$ triangle decorations separated by $\left \lfloor{N/m}\right \rfloor$ edges. This value jumps when $N\%m = 0$, resulting in the observed discontinuous changes at these values when $\aff_0/N = 0.5$.}
\label{fig:r-t}
\end{figure}

First we test the accuracy of the coarse-grained and theoretical approximations for a fixed network topology. In Fig.~\ref{fig:r-t} we show results for networks with a large cycle of size $N = 100$ with a single triangle decoration as a function of $\mu$, and with $m$ evenly spaced triangle decorations as a function of $m$. All of the rates in the large cycle are set to $k^- = 1$ and $k^+ = \exp(\aff_0/N)$ where $\aff_0$ is the affinity in a network with no decorations, and the rates in the decorations are $a = k^-$ and $b = k^+$. When $m = 1$, $T/\R_{CG}$ approaches $T/\R_{exact}$ as the affinity increases, with perfect agreement in the limit of very high affinity ($k^-/k^+ = \exp(-10) \approx 0$). Note that in this limit the effective reverse hopping rates along edges representing coarse-grained decorations are not suppressed; rather they are enhanced (Fig.~\ref{fig:effrates}) and $ \eta^-/\eta^+> 1$ (Fig.~\ref{fig:effrates}f), so that even in the limit $k^-/k^+ \to 0$ it is non-trivial to predict the period.  The agreement between $T/\R_{th}$ and $T/\R_{CG}$ is also excellent. The net effect is convergence between exact values and theoretical predictions for $T$ and $\R$ with increasing affinity.

The distance between decorations is given by $\left \lfloor{N/m}\right \rfloor$. At low affinity, this results in discontinuous jumps in the values of $T/\R_{exact}$ and $T/\R_{CG}$ (yellow lines in right column of Fig.~\ref{fig:r-t}) at values of $m$ where $N$ is an integer multiple of $m$, because the distance between the decorations is important.  At high affinity the distance no longer matters, as predicted by our theory, and the CG and exact lines become smooth and ultimately match the theory prediction.

\subsection{Networks with randomly distributed decorations}

\begin{figure}
\centering
\includegraphics[width=\linewidth]{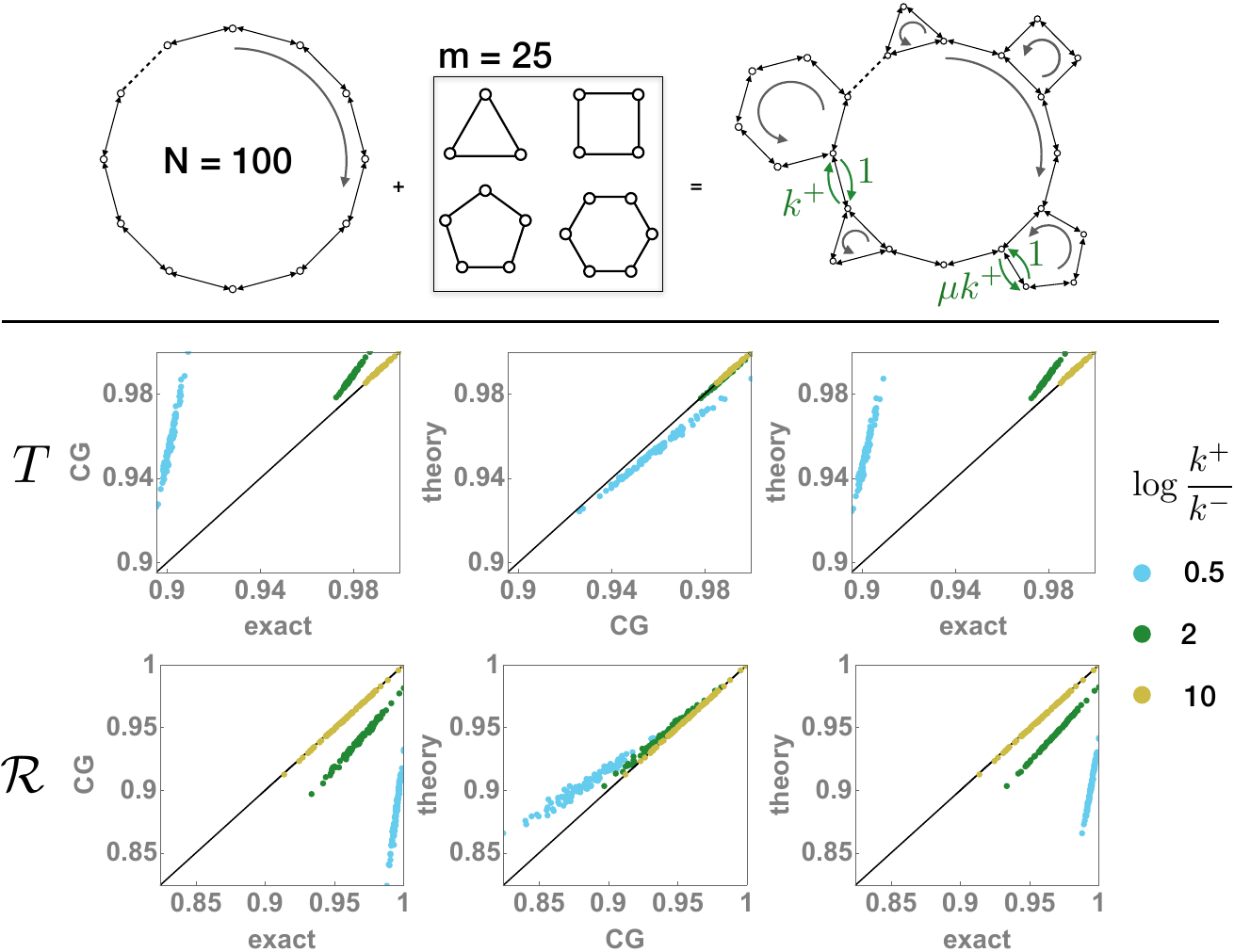}
 \caption{Period $T$ and number of coherent oscillations $\R$ in networks with $N = 100$ states in the main cycle and $m = 25$ decorations whose shape and location are randomly generated, with the constraint that no vertex is connected to more than three neighbors, i.e. the decorations are separated by at least one edge. The rates are chosen as in Fig.~\ref{fig:r-t}, except that we chose $\mu = 0.05$ since the effective rates for larger decorations diverge at decreasing values of $\mu$ (see Table~\ref{tab:rates}).  Each point in the scatter plots represents a specific realization of quenched disorder of shapes and locations. As $\aff_0/N = \log(k^+/k^-)$ increases, the exact, coarse-grained and theory values converge. All values are normalized by the largest value in the data on each scatter plot so that data for different affinities can be shown on the same plot.}
\label{fig:r-t-random}
\end{figure}

The success of our theory in predicting time scales in Fig.~\ref{fig:r-t} suggests that networks with many different arrangements of the same set of decorations, or of similar sets drawn from a common distribution, can have the same values of $T$ and $\R$. We now introduce disorder by fixing the number of decorations $m = 25$ and the value of $\mu = 0.05$ and selecting the shapes and locations of decorations in the network randomly (with all other parameters the same as in Fig.~\ref{fig:r-t}). Because our coarse-graining scheme takes into account the edges CW and CCW from the decoration as shown in Fig.~\ref{fig:schematic}b, we do not expect it to work well when two decorations are connected to the same vertex, and we constrain the random locations so that this does not happen (i.e. decorations are separated by at least one edge). The decorations have 3 - 6 sides. Scatter plots in which each point represents one realization of the quenched shape and location disorder are shown in Fig.~\ref{fig:r-t-random}. The results show that in these disordered networks, the exact, CG, and theory results converge at high affinity, confirming that the arrangement of decorations is unimportant at high affinity.

\subsection{Combining rate disorder and topological disorder}

\begin{figure}
\centering
\includegraphics[width=\linewidth]{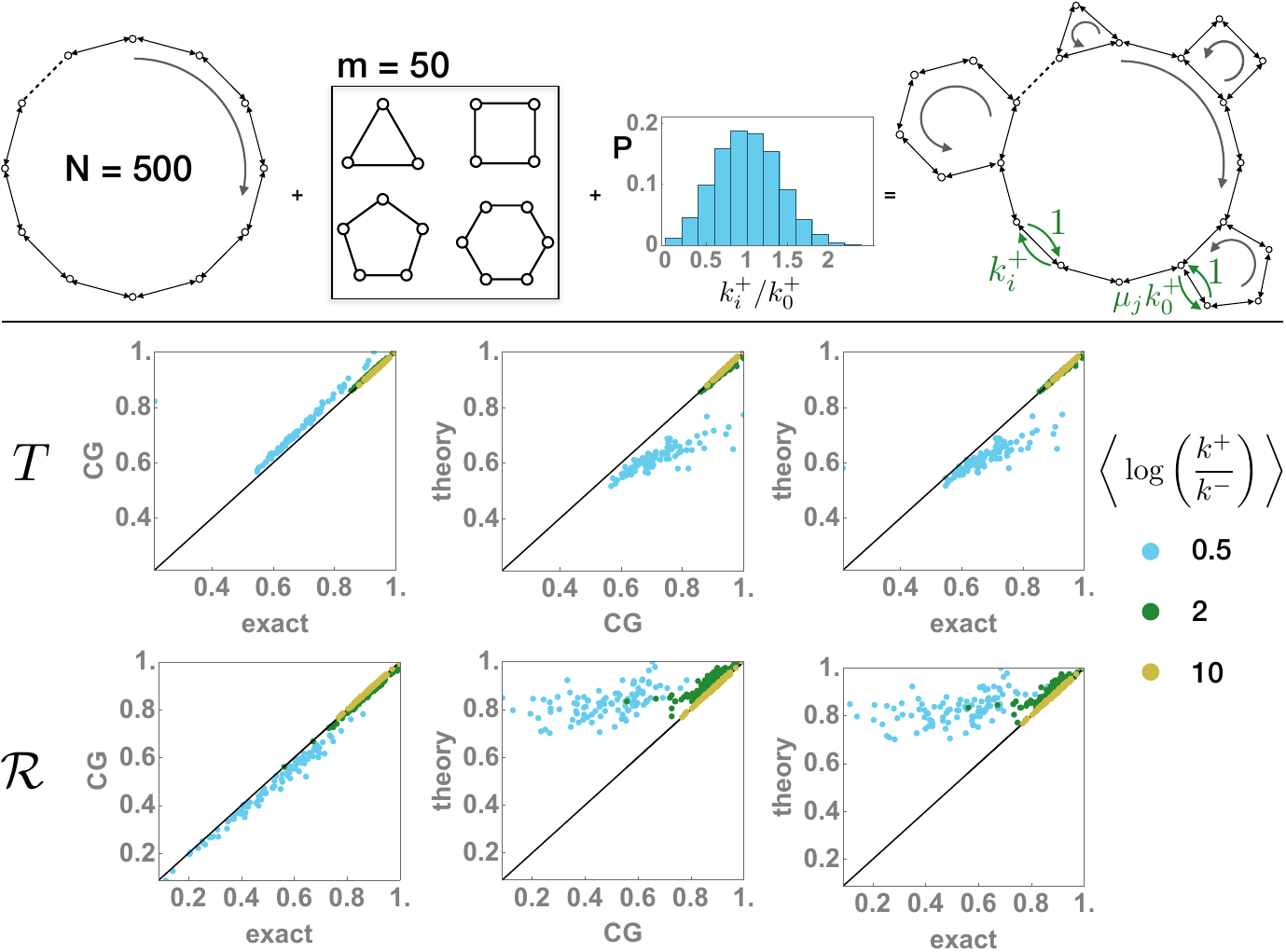}
 \caption{Period $T$ and number of coherent oscillations $\R$ in networks with $N = 500$ states in the main cycle and $m = 50$ decorations. Here, disorder in the rates of the main cycle as well as disorder in the topology of the network are considered. The shapes and locations of the decorations as well as many of the rates are randomly generated as described in the text. Each point in the scatter plots represents a specific realization of quenched disorder of shapes, locations, and rates. As $\aff_0/N = \langle k^+/k^-\rangle$ increases, the exact, coarse-grained and theory values converge. All values are normalized by the largest value in the data on each scatter plot so that data for different affinities can be shown on the same plot.}
\label{fig:r-t-mix}
\end{figure}

Finally, we test how our theory performs when disorder in the rates, explored in Ref.~\citenum{DelJunco2018b}, is combined with random network topology (Fig.~\ref{fig:r-t-mix}). We now generate networks with large cycles of size $N = 500$ and $m = 50$ decorations with random shapes and random locations. The locations are again constrained so that no two decorations are connected to the same vertex. The CCW rates in the large cycle are set to $k^- = 1$, and the CW rates that are not part of the coarse-grained motif (i.e., are not connected to a vertex which is part of a decoration) are then chosen randomly from a Gaussian distribution with mean $k^+_0 = \mathcal A_0/N$ and standard deviation $0.4k^+_0$ and a lower cutoff at $0.1k^+_0$. We set the rates in the motifs to the {\it cis} configuration: $a = 1, b = k^+_0$, and we introduced disorder in the probability of entering the decorations by choosing $\mu$ randomly from a uniform distribution $[0, 0.95\mu_{max}]$, where $\mu_{max}$ is the value of $\mu$ at which the effective rates in Table~\ref{tab:rates} diverge, which depends on the size of the decoration. Once again, the agreement between both levels of approximation and exact results for $T$ and $\R$ is excellent in the limit of high affinity ($\aff_0/N  = \exp(10)$), with very good agreement already at moderate values of the affinity ($\aff_0/N  = \exp(2)$). 

\section{Discussion}\label{disc}

\begin{figure}
\centering
\includegraphics[width=0.9\linewidth]{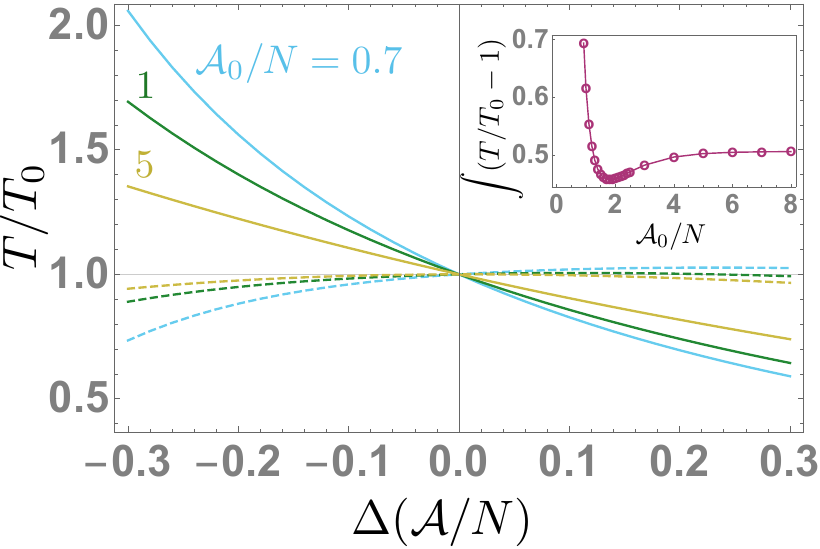}
 \caption{Compensating for changes in period by tuning the parameter $\mu$ as a function of $\Delta\aff$ in a network with a large cycle of $N = 100$ states with $m = 20$ triangle decorations symmetrically distributed on the network, with the {\it cis} configuration ($a = k^- = 1, b = k^+ = \exp(\aff/N) = \exp((\aff_0 + \Delta\aff)/N)$). $T_0$ is the period at $\aff_0/N$ and $\mu_0 = 0.5$. Dashed lines are the period as the affinity is changed with $\mu = \mu_0$; solid lines are the period when $\mu$ is changed as a function of the affinity: $\mu =\mu_0 + \Delta\mu$ with $\Delta\mu = \kappa_{comp}\Delta\aff$, as described in the text. This linear compensation mechanism is effective over a wide range of affinities $\aff_0$, compensating for changes in the period of $\sim 50\%$. Compensation occurs because by increasing (decreasing) $\mu$ the system is encouraged to spend more (less) time in the triangle decorations as the affinity is increased (decreased). Inset: The integrated deviation from perfect compensation ($T/T_0 = 1$) over the range $\Delta\aff \in [-0.3, 0.3]$ as a function of the affinity in the unperturbed network.  The mechanism is most effective when $\aff_0/N \approx 2$ or greater.}
\label{fig:compensation}
\end{figure}


The results of Fig.~\ref{fig:r-t-mix} show that the time scales of an oscillator with multiple cycles and randomly distributed rates does not depend on the arrangement of these rates and cycles.  As a result, these observables can be accurately predicted from our theory with information about the probability distributions of the rates and decorations, and notably without information about the spatial arrangement of the specific network. This extends the conclusions of Ref.~\citenum{DelJunco2018b} to the case of network topology. The motivation for studying multiple cycles is that the small cycles can represent deviations from or noise in the oscillator's average limit cycle~\cite{Pittayakanchit2018, Marsland2019}. The quenched disorder in Figs.~\ref{fig:r-t-random} and \ref{fig:r-t-mix} are meant to represent different realizations of the pathways sampled by the oscillator over time, or by multiple copies of the same oscillator, e.g. in different cells. In this context, our results show how an oscillator whose sampled paths and rates are fluctuating over time can use a high chemical affinity to maintain a predictable and robust period.



So far we have considered disorder that kept the average value of the rates and affinity constant. We now briefly turn our attention to global fluctuations in the rates that result in a change in the affinity; for example, this could be due a shift in the overall ATP to ADP ratio in a cell caused by a change in light levels or a change in temperature. Biochemical oscillators often have the ability to maintain a constant period in the presence of these changes, a feature known as input compensation~\cite{Johnson2011, Francois2012, Paijmans2017}. For a given network topology and arrangement of relative rate magnitudes, changing the affinity effectively multiplies all of the rates by a constant since $k^+/k^- \propto \exp(\aff/N) = \exp(\aff_0/N) \exp(\Delta\aff/N)$, where now $\aff_0$ refers to the reference or unperturbed value of the affinity in the main cycle of a decorated network. Any change in the affinity therefore results in a change in the period. However, if the rates or the decorations in the network are allowed to vary in a manner that is coupled to the change in affinity, the oscillator may be able to keep a constant period.  Specifically, if the current increases on the network in response to an increase in the affinity, the system can increase the path length by increasing the probability of entering and remaining in decorations.  This mechanism for compensation is a stochastic version of one that has previously been explored in deterministic limit cycles by several authors~\cite{Francois2012, Hatakeyama2015}: if an input changes the angular velocity of the limit cycle, the radius of the limit cycle must also change in response to the input in order to maintain a constant period. 

We illustrate this compensation mechanism in Fig.~\ref{fig:compensation}. We choose the rates in the network as in Fig.~\ref{fig:r-t}: $a = k^- = 1$, $b = k^+ = \exp(\aff/N)$. First we hold $\mu$ fixed and vary the affinity so that the rates become $b = k^+ = \exp((\aff_0 + \Delta\aff)/N)$, and show that the period changes significantly with small changes in the affinity (solid lines in Fig.~\ref{fig:compensation}) - for instance, changing the affinity from $\aff_0/N = 5$ to $\aff/N = 5.3$ shortens the period by 25\%. Then, we allow the value of $\mu$ to be appropriately coupled to the affinity and show that these changes in the period can be compensated for, reducing the change to less than 5\%. The parameter $\mu$ controls the probability of accessing the smaller secondary cycles. Hence, the parameter $\mu$ effectively controls the size of sampled orbits in our networks. 

In order to choose how $\mu$ should depend on the affinity, we consider the Taylor expansion of the period as a function of $\aff$ and $\mu$:
\begin{align}
T(\aff, \mu) = &T(\aff_0, \mu_0)  \\
&+  \left(\frac{\partial T(\aff_0, \mu_0) }{\partial \mu}\right) \Delta\mu +  \left(\frac{\partial T(\aff_0, \mu_0) }{\partial \aff}\right) \Delta\aff \notag \\
& + \mathcal O((\Delta\mu)^2, (\Delta\aff)^2, \Delta\mu\Delta\aff) + \cdots .\notag
\label{eq:texpand}
\end{align}
Perfect compensation then requires $T(\aff, \mu) = T(\aff_0, \mu_0)$ or 
\begin{equation}
0 =  \left(\frac{\partial T(\aff_0, \mu_0) }{\partial \mu}\right) \Delta\mu +  \left(\frac{\partial T(\aff_0, \mu_0) }{\partial \aff}\right) \Delta\aff + \cdots
\label{eq:texpand}
\end{equation}
for all $\Delta\aff, \Delta\mu$.
In general this leads to a very complicated $\Delta\mu$ that is a function of $\Delta\aff$ with as many parameters as the Taylor expansion has terms. However, in Fig.~\ref{fig:compensation} we find numerically that over large changes in the period, $T$ it is in fact a linear function of $\Delta\aff$ and $\Delta\mu$, so that we can achieve compensation just by setting 
\begin{align}
\Delta\mu &= - \left[\left(\frac{\partial T(\aff_0, \mu_0) }{\partial \aff}\right)\bigg/\left(\frac{\partial T(\aff_0, \mu_0) }{\partial \mu}\right) \right]\Delta\aff\\
 &\equiv \kappa_{comp} \Delta\aff.
\end{align}

In the inset in Fig.~\ref{fig:compensation}, we see that this `linear compensation' mechanism works best above a minimum value of the affinity around $\aff_0/N = 2$, indicating that a high chemical affinity can support simple mechanisms for compensation.
Indeed, using a linear approximation of our theory in Eqs.~\ref{eq:phinew} - \ref{eq:gamma}, we find that as long as the effective rates in the coarse-grained link are proportional to $k^+$ (as they are in our case; see Table~\ref{tab:rates}), all second-order and higher terms in Eq.~\ref{eq:texpand} vanish at high affinity and for large main cycle size $N$. High affinity therefore makes it easy to design (or evolve) a network of this kind with compensation, since only one parameter needs to be set, which is easily computed from the unperturbed ($\mu_0, \aff_0$) network.

Here we have illustrated compensation using $\mu$ for simplicity since it is a continuous variable. However, the number of decorations or the size of the decorations could also be used to adjust the period, since these all affect the path length of an oscillation, or alternately, the amount the time the system spends effectuating futile cycles in decorations. 

\section{Conclusion}\label{conc}

In this paper we presented an analytical theory for computing the period of oscillations in Markov models consisting of one large cycle of size $N$ decorated with smaller secondary cycles that are driven out of equilibrium by an affinity $\mathcal A$ (Fig.~\ref{fig:schematic}). First, we mapped the decorations on to single links that retain the mean and variance of the first passage time across the decoration. Performing this procedure for all of the decorations in the network yields a single-cycle network for which we have previously derived analytical expressions for the period and coherence of oscillations. Importantly, these analytical expressions take as input the rates along each edge in the network, but do not know about their relative placement. Numerical calculations of the period at high affinity agree well with this analytical prediction (Figs.~\ref{fig:r-t} - \ref{fig:r-t-mix}). Our main result is that the ability of our theory to accurately predict the period and coherence implies that high energy dissipation makes these observables insensitive to many parameters; specifically, the arrangement of the cycles and rates in the network. As a result, oscillators represented by the models studied here can have time scales that are robust to fluctuations in rates and topology. Finally, we showed how multi-cycle network topologies can also be exploited to achieve compensation to changes in affinity, by tuning the amount of time that the system spends in the decorations.

\begin{acknowledgments}

Thanks to Kabir Husain for generative discussions and for explaining the method to calculate first passage time distributions, and to Mike Rust for helpful discussions. We wish to acknowledge constructive comments from anonymous reviewers of Ref.~\citenum{DelJunco2018b}, which partially motivated this work, and specifically the reviewer who suggested the scatter plot presentation of data used in Figs.~\ref{fig:r-t} - \ref{fig:r-t-mix}. CdJ acknowledges the support of the Natural Sciences and Engineering Research Council of Canada (NSERC). CdJ a \'et\'e financ\'ee par le Conseil de recherches en sciences naturelles et en g\'enie du Canada (CRSNG). This work was partially supported by the University of Chicago Materials Research Science and Engineering Center (MRSEC), which is funded by the National Science Foundation under award number DMR-1420709.  SV also acknowledges support from the Sloan Fellowship and the University of Chicago.

\end{acknowledgments}


%

\appendix

\section{Theory from Ref.~\citenum{DelJunco2018b}}\label{app:theory}

Our theory in Ref.~\citenum{DelJunco2018b} uses a transfer matrix formulation of the eigenvalue equation for the transition rate matrix of a single-cycle network of size $N$ where all but one set of rates are the same:
\begin{equation}
\begin{bmatrix}
f_1 \\ f_2
\end{bmatrix}
=
\MA\MB^{N-1}
\begin{bmatrix}
f_1 \\ f_2
\end{bmatrix}
\label{eq:transfer0}
\end{equation}
where $f_i$ are eigenvector elements, $\MB$ is a transfer matrix mapping eigenvector magnitudes about links with `uniform rates' $k^\pm$, and $\MA$ is a transfer matrix mapping eigenvector magnitudes about the link with `defect rates' $h^\pm$.  $\MA$ and $\MB$ are functions of the eigenvalue $\phi$ of the transition rate matrix. By Eq.~\ref{eq:transfer0}, $\MA\MB^{N-1}$ must have an eigenvalue of one. We postulate $\phi = \phi^{(0)} + C \gamma$ with $C$ a constant given in Eq.~\ref{eq:c}. Therefore, Eq.~\ref{eq:transfer0} is a self-consistent equation for $\gamma$ which we solve as described in Ref.~\citenum{DelJunco2018b} to obtain Eq.~\ref{eq:gamma}. Eq.~\ref{eq:transfer0} is easily extended to cases where there is more than one set of defect rates; further details can be found in Ref.~\citenum{DelJunco2018b}. 

In Ref.~\citenum{DelJunco2018b} we approximated the product of transfer matrices as:
\begin{align}
\MA  \MB^{N-1} &\approx \MA\left[ \left(\beta_1^{(1)}\right)^{N-1} \MX_1^{(0)} + \left(\beta_2^{(1)}\right)^{N-1} \MX_2^{(0)} \right] \\
& \approx
\left(\beta_1^{(1)}\right)^{N-1}\MA \MX_1^{(0)}
\label{eq:transfer}
\end{align}
where $\beta_1^{(1)} = \exp(2\pi i/N) (1 + \gamma) =\beta_1^{(0)}(1 + \gamma)  $ and $\beta_2^{(1)} = (k^-/k^+)\exp(2\pi i/N) (1 + \gamma) = \beta_2^{(0)}(1 + \gamma) $ are first-order perturbed eigenvalues of $\MB$, and $\MX_i^{(0)} = |i^{(0)}\rangle \langle i^{(0)} |$ is the outer product of the $i$th unperturbed eigenvectors. In the second line we have assumed high affinity : $k^-/k^+ \ll 1$. The $i$th left and right eigenvectors of $\MB$ are given by:
\begin{align}
\langle i | =  \{ -1/\beta_j, 1\}/c_i^2 && | i \rangle = \{ \beta_i, 1\}/c_i^2
\end{align}
where $c_i$ is a normalization constant. Therefore, by using $\MX_1^{(0)}$, our theory ignored important terms containing $\gamma$. In cases where $k^-/k^+ \ll 1$ and $h^-/h^+ \ll 1$, we find that these terms cancel and our theory works with $\MX_1^{(0)}$, explaining the success of our theory in predicting timescales in Ref.~\citenum{DelJunco2018b}. However, in the coarse-grained networks studied in this paper, specifically for decorations with the $cis$ configuration, we often have $h^-/h^+ > 1$.  We therefore replace 
\begin{align}
\MX_1^{(0)} & =
\frac{1}{c_1^2}   \begin{bmatrix} 
  -\beta_1^{(0)}/\beta_2^{(0)} & \beta_1^{(0)} \\
 -1/\beta_2^{(0)} & 1
\end{bmatrix}\\
& \to
\MX_1^{(1)} =
\frac{1}{c_1^2}   \begin{bmatrix} 
  -\beta_1^{(1)}/\beta_2^{(1)} & \beta_1^{(1)} \\
 -1/\beta_2^{(1)} & 1
\end{bmatrix}
\end{align}
in Eq.~\ref{eq:transfer}, and proceed with the calculation as described in the Supplementary Material of Ref.~\citenum{DelJunco2018b}, to obtain Eqs.~\ref{eq:phinew} - \ref{eq:gamma}, where
\begin{widetext}
\begin{align}
\zeta_j &=  \frac{{h_j^-} {k^+}+{h_j^+} {k^-}-{k^-} {k^+} + 2 \gamma  {h_j^+} {k^-} +\gamma ^2 \left({h_j^+} {k^-} + {k^-} {k^+} \right) + (\gamma +1) {k^+} e^{\frac{2 i \pi }{N}} (-{h_j^-}-{h_j^+}+{k^-}+{k^+})-\left((\gamma +1) {k^+} e^{\frac{2 i \pi }{N}}\right)^2}{(\gamma +1) {h_j^+} \left({k^-}-{k^+} e^{\frac{4 i \pi }{N}}\right)} \label{eq:zeta} \\
C &=  c_1^2 k^- e^{-2\pi i/N} \label{eq:c} \\
c_1^2 & = 1-(k^+/k^-)e^{4\pi i/N}. \label{eq:c1sq}
\end{align}
\end{widetext}

\section{Calculating the first passage time distribution}\label{app:fpt}

The method for calculating the first passage time between two states is to sum over all of the paths of all lengths connecting the two states.  First we write down the FPT distribution between two connected states $1$ and $2$. If the system enters state $1$ at time $t_0$, the probability that it hops to state $j$ at time $t_1 = t_0 + \mu  t$ is:

\begin{equation}
Q_{12}(\mu  t)  = P_{12}(\mu  t) \prod_{i \neq 1, 2}\left( 1 - \int_0^{\mu  t} dt P_{1i}(t) \right)
\end{equation}
Where $P_{12}(t) = k_{12}\exp(-k_{12}t)$ is the waiting time distribution for hopping from state 1 to 2. The first term is the probability of hopping at exactly time $t_1$, while the term in parentheses is the probability that the system has not hopped to any other state in the meantime. The net waiting time distribution out of state 1 is just the sum over connected states: $\sum_j Q_{1j}$. 

The probability of observing a particular trajectory with transitions $\{1\to 2, 2 \to 3, \cdots, n-1\to n\}$ occurring at times $\{t_1, t_2, t_3, \cdots, t_{n-1}\}$ is:
\begin{equation}
Q_{12}(t_1)Q_{23}(t_2-t_1)\cdots Q_{n-1, n}(t_{n-1} - t_{n-2}).
\end{equation}
Note that state 1 is the first state that the system visits, it is not necessarily a state with a fixed label. For example, state 1 and state 3 could both be the same state $i$, if the system's trajectory is $i \to j \to i$. State $n$ is the only state that can be visited only once, since it is an absorbing state. 

To obtain the first passage time distribution $F(n, t_{n-1} | 1, 0)$ we sum over all trajectories that start at state 1 at time $t_0$ and arrive, for the first time, at state $n$ at time $t_{n-1}$. We integrate over all possible combinations of transition times $\{t_1, \cdots t_{n-2}\}$ constrained such that $t_{n-1}$ is fixed, and sum over all paths that the system can take:
\begin{align}
&F(n, t_{n-1}| 1, 0) =  \\
&\sum_{paths} \int dt_1 \cdots dt_{n-2} Q_{12}(t_1) \cdots Q_{n-1, n}(t_{n-1} - t_{n-2}). \notag
\end{align}
We can turn this convolution into a product by taking the Laplace transform:
\begin{equation}
\hat Q_{ij} = \int_0^\infty dt e^{-st} P_{ij} (t)  \prod_{k \neq i, j}\left( 1 - \int_0^{t} dt' P_{ik}(t') \right)
\end{equation}
so that we have
\begin{equation}
\hat F(n, t_{n-1}| 1, 0) =\sum_{paths} \hat Q_{12} \times \hat Q_{13}\times  \cdots\times \hat Q_{n-1, n}.
\end{equation}
Since each transition in our Markov model is a Poisson process, we plug in an exponential form for $P$, giving:
\begin{align}
&\hat Q_{ij} (s) = \\
&k_{ij} \int_0^\infty dt e^{-st} e^{-k_{ij}t} \prod_{k \neq i, j}\left( 1 - k_{ik}\int_0^{t} dt' e^{-k_{ik}t'} \right) \\
& =  \frac{k_{ij}}{s+\sum_{k \neq i}k_{ik}}.
\end{align}
To sum over paths we will construct a matrix $\mathbf K$ with elements 
\begin{equation}
K_{ij} = \left\{ \begin{matrix}  \hat Q_{ij} (s) & \text{if states i and j are connected}  \\   0 &  \text{otherwise} \end{matrix} \right.
\end{equation}
$K_{1n}$ then gives us the waiting time distributions for all paths of length 1 from state 1 to $n$. $[K^2]_{1n}$ gives us paths of length 2, and $[K^m]_{1n}$ gives all paths of length $m$. Summing,
\begin{align}
\hat F(n, t_{n-1}| 1, 0) &= 1 + K_{1n} + [K^2]_{1n} + \cdots  \\
&= \sum_{m=0}^\infty [K^m]_{1n} \\
&= [(\mathbbm 1  - \mathbf K)^{-1}]_{1n}.
\end{align}
All elements of the matrix $\mathbf K$ are strictly less than 1 since $s$ is always positive, so the Frobenius norm of the matrix $\lim_{m\to \infty}\mathbf K^m <1$ and the series converges. We can then either invert the Laplace transform to obtain the FPT distribution $F(t)$, or if that's not tractable, we can obtain the moments of the distribution using Eq.~\ref{eq:genfunc}, which is what we do in this paper.

\begin{widetext}

\section{Effective rates for a triangle decoration}\label{app:eff-rates}

\begin{align}
\lambda \eta^+ &= ({k^-}+{k^+})^2 \left(a^2 \mu  +{k^+} (a+b)\right)^2\\
\lambda \eta^-  &=  \mu^3 a^3 b  {k^+}^2 \notag \\ 
&+  \mu  ^2 a b \left[a b ({k^-}+{k^+})^2+a {k^+} \left(-{k^-}^2-2 {k^-} {k^+}+{k^+}^2\right)+{k^+}^3 (b+{k^-}+{k^+})\right] \notag \\ 
&+\mu \Bigl[ a^3 {k^-} ({k^-}+{k^+})^2-a^2 ({k^-}+{k^+})^2 ({k^+} ({k^-}+{k^+})-b {k^-}) \notag \\
& +a {k^+} \left(b^2 ({k^-}+{k^+})^2-b {k^-} {k^+} ({k^-}+2 {k^+})+{k^+}^2 ({k^-}+{k^+})^2\right) \notag \\ 
&+b {k^+} \left(b^2 ({k^-}+{k^+})^2+b {k^+}^3+{k^+}^3 ({k^-}+{k^+})\right)\Bigr] \notag \\ 
& +{k^-} {k^+} (a+b)^2 ({k^-}+{k^+})^2 \\
\lambda &= \mu^3 a^3 b  {k^-}\notag \\
&+a \mu  ^2 \left[a^2 ({k^-}+{k^+})^2+a b \left({k^-}^2+4 {k^-} {k^+}+{k^+}^2\right)+b {k^+} (b {k^-}-{k^+} ({k^-}+{k^+}))\right] \notag \\ 
&+\mu   \Bigl[ a^3 ({k^-}+{k^+})^2+a^2 ({k^-}+{k^+})^2 (b+{k^-}+2 {k^+}) \notag \\
&-a {k^+} \left({k^+} ({k^-}+{k^+})^2-b (2 {k^-}+{k^+}) ({k^-}+2 {k^+})\right)+b {k^+}^2 (b {k^-}-{k^+} ({k^-}+{k^+}))\Bigr] \notag \\ 
&+{k^+} (a+b)^2 ({k^-}+{k^+})^2
\end{align}
\end{widetext}

\end{document}